\documentclass[aps,pra,twocolumn,showpacs]{revtex4}
\usepackage{amsmath,amssymb,graphicx,bbold}
\usepackage{graphicx}  % needed for figures
\usepackage{dcolumn}   % needed for some tables
\usepackage{bm}        % for math
\usepackage{color}
\usepackage{epsfig}
\usepackage{epstopdf}
\usepackage{subfigure}
\providecommand{\abs}[1]{\left\lvert#1\right\rvert}

\providecommand{\ket}[1]{\lvert #1 \rangle}
\providecommand{\braket}[2]{\langle #1 \rvert #2 \rangle}

\begin{document}

\title{Intrinsic bounds of a two-qudit random evolution}
 
\author{A.~Z.~Khoury$^1$, A.~M.~Souza$^2$, L.~E.~Oxman$^1$, I.~Roditi$^2$, R.~S.~Sarthour$^2$, I.~S.~Oliveira$^2$}

\affiliation{
1- Instituto de F\'{\i}sica, Universidade Federal Fluminense, 24210-346, Niter\'oi - RJ, Brazil.\\
2- Centro Brasileiro de Pesquisas F\'{\i}sicas, Rua Dr. Xavier Sigaud 150, 22290-180, Rio de Janeiro - RJ, Brazil.}

%\date{\today}
 
\begin{abstract}
We investigate entangled qudits evolving under random, local $SU(d)$ operations and demonstrate that this evolution 
is constrained by intrinsic bounds, showing robust features of two-qudit entangled states that can be useful for fault 
tolerant implementations of phase gates. Our analytical results are supported by numerical simulations and confirmed 
by experiments on liquid-state nuclear magnetic resonance qubits.
\end{abstract}
\vskip2pc 
 
\maketitle

The development of quantum technologies is challenged by the unavoidable action of the environment 
and uncontrollable experimental imperfections. These effects cause errors in quantum 
algorithms and limit the scalability of quantum devices. Strategies to isolate the physical systems 
used for quantum information tasks and to identify some of their features immune to those undesired  
effects are in course. The use of engineered reservoirs \cite{kwiat,aiello,global}, 
decoherence free subspaces \cite{decfree}, 
geometrical phases \cite{vedral}, topologically protected systems \cite{faulttolerant} and dynamical 
decoupling (see for example \cite{dd1,dd2,dd3,dd4}) have been considered as potential 
means for robust quantum computation. The role of entanglement in the geometric phase acquired by entangled 
quantum systems has motivated a great deal of interesting research works \cite{sjoqvist,sjoqvist2}.
In particular, it has been theoretically predicted \cite{milman,milman2} and experimentally demonstrated 
\cite{topoluff,nmr} that the geometric phase acquired by maximally entangled qubits is discrete, restricted 
to integer multiples of $\pi\,$. Later, fractional phases were predicted for maximally entangled systems with 
arbitrary dimension $d$ (\textit{qudits}) \cite{fracuff,fracuff2,fracuffufmg,fracuffufmg2} and for multiple 
qubits \cite{multiqubit,multiqubit2}. 
Quantum algorithms employing qudits have also been considered in the literature \cite{ashok,bullock,munro}. 

In this work we demonstrate an intriguing feature of entanglement regarding the evolution of a two-qudit 
state under 
arbitrary local $SU(d)$ transformations. The inner product (overlap) between the transformed and the initial states 
is shown to be confined within a nontrivial boundary in the complex plane. This boundary is deduced 
analytically, confirmed by numerical simulations with random local $SU(d)$ transformations and, furthermore, 
experimentally verified with nuclear magnetic resonance.  

Let us consider a two-qudit quantum state evolution 
\begin{eqnarray}
\ket{\psi(t)} = \sum_{i,j=1}^{d} M_{ij}(t) \ket{ij}\;,
\end{eqnarray}
where $M$ is the coefficient matrix of the quantum state expansion in the computational 
basis $\{\ket{ij}\}\,$. It will be useful for our purposes, to write the coefficient 
matrix in its polar decomposition:
\begin{eqnarray}
M(t) = e^{i\theta(t)}\,Q(t)\,S(t)\;,
\end{eqnarray}
where three sectors can be identified: 
\begin{itemize}
	\item The $U(1)$ sector represented by the overall phase factor $e^{i\theta(t)}\,$. 
	\item The $SU(d)$ sector given by $S(t)\in SU(d)\,$. 
	\item The Hermitian sector $Q(t)=Q^{\dagger}(t)\,$.
\end{itemize}
Under local unitary transformations, each sector follows an independent time evolution. 
Indeed, let $U_A(t)=e^{i\theta_A(t)}\bar{U}_A(t)$ and $U_B(t)=e^{i\theta_B(t)}\bar{U}_B(t)$ 
be the local unitary transformations applied to qudits $A$ and $B\,$,
with $\bar{U}_{A}(t)$ and $\bar{U}_{B}(t) \in SU(d)\,$.
In this case the evolution in each sector is given by
\begin{eqnarray}
\theta(t) &=& \theta(0) + \theta_{A}(t) + \theta_{B}(t)\;,
\nonumber\\
Q(t) &=& \bar{U}_{A}(t)\,Q(0)\,\bar{U}_{A}^{\dagger}\;,
\label{sectors}\\
S(t) &=& \bar{U}_{A}(t)\,S(0)\,\bar{U}_{B}^{T}\;.
\nonumber
\end{eqnarray}
It will be convenient to adopt the basis leading to the Schmidt decomposition 
of the initial state, 
so that $M_{ij}(0) = M_{jj}(0)\,\delta_{ij}$ and $M_{jj}(0)\in\mathbb{R}\,$.
In this case,  $\theta(0)=0\,$ and $S(0) = \mathbb{1}\,$. Moreover, we shall 
assume that the two qudits are locally operated with $SU(d)$ transformations 
($\theta_{A} = \theta_{B} = 0$), making the $U(1)$ sector stationary. Note 
that local $SU(d)$ transformations are naturally realized in 
spin systems interacting with an external magnetic field, since the 
energies of the spin eigenstates are symmetrically shifted in this case. 
As we will see, this makes nuclear magnetic resonance (NMR) the ideal 
platform for the experimental investigation.

\begin{figure}[h!]
	\includegraphics[scale=0.6]{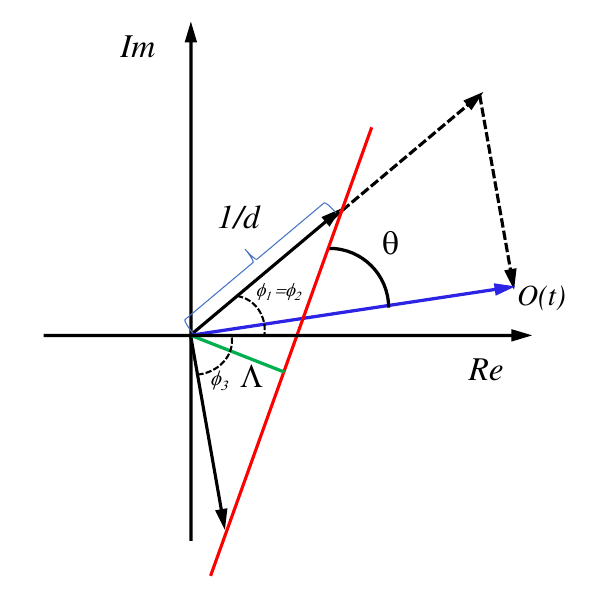}
	\caption{\label{fig:fasores-max-entang} Graphic representation of the $SU(3)$ eigenvalues and the maximum 
		state overlap, showing the geometric meaning of the Lagrange variables.}
\end{figure}

The intrinsic bounds imposed on entangled qudits are more pronounced for 
maximally entangled states, as we will see shortly. 
In this case, $Q(0)=\mathbb{1}/\sqrt{d}$ and the 
Hermitian sector remains stationary for arbitrary local transformations. Then, 
the two-qudit state evolution will be restricted to the $SU(d)$ sector of the 
coefficient matrix and the time dependent overlap between the evolved and the 
initial states will be 
\begin{eqnarray}
O(t)=\braket{\psi(0)}{\psi(t)} = Tr\left[M^\dagger(0)\,M(t)\right] = 
\frac{Tr\left[S(t)\right]}{d}\;.
\nonumber\\
\end{eqnarray}
We now demonstrate that this overlap remains restricted to a confined area of the 
complex plane, whose perimeter depends on the dimension of the qudits. Since $S(t)\in SU(d)$ 
($\det S(t) = 1$), its eigenvalues are phase factors $\{e^{i\phi_j(t)}\}$ ($j=1\dots d)$, 
constrained by the condition 
\begin{eqnarray}
F(\{\phi_j\}) = \sum_{j=1}^d \phi_j = 0\;.
%\Rightarrow\;\phi_d = -\sum_{j=1}^{d-1} \phi_j. 
\label{phid}
\end{eqnarray}
In terms of these phase factors, the time dependent overlap is given by 
\begin{eqnarray}
O(t) = \frac{1}{d} \sum_{j=1}^d e^{i\phi_j(t)} = R \,e^{i\Phi}\;, 
\label{Ot}
\end{eqnarray}
where $R$ is the overlap absolute value and $\Phi$ its phase. In order to find the contour of 
the overlap boundary in the complex plane, we need to determine $R_{max}(\Phi)\,$, 
the extrema of $R$ for each phase $\Phi\,$. For a given eigenvalue configuration $\{\phi_j\}\,$, 
the overlap absolute value can be deduced from the real part of Eq. (\ref{Ot}) as 
\begin{eqnarray}
R (\{\phi_j\}) = \frac{1}{d}\sum_{j=1}^{d} \cos\left(\phi_j - \Phi\right) \;. 
\label{rophi0}
\end{eqnarray}
At the same time, a second constraint can be immediately derived from the imaginary part 
\begin{eqnarray}
G(\{\phi_j\}) = \frac{1}{d}\sum_{j=1}^{d} \sin\left(\phi_j - \Phi\right) = 0\;. 
\label{pphi}
\end{eqnarray}
We are now able to find $R_{max}(\Phi)$ under these two constraints using Lagrange multipliers. 
For this end, we define
\begin{eqnarray}
L (\{\phi_j\},\lambda_f,\lambda_g) = R (\{\phi_j\}) + \lambda_f\,F(\{\phi_j\}) + \lambda_g\,G(\{\phi_j\}) \;,
\nonumber\\ 
\label{L}
\end{eqnarray}
where $\lambda_f$ and $\lambda_g$ are Lagrange multipliers. 
The constrained extrema of $R$ are obtained from the solutions of 
\begin{eqnarray}
\frac{\partial L}{\partial\phi_k} &=& \frac{-\sin\left(\phi_k - \Phi\right) + \lambda_g\,\cos\left(\phi_k - \Phi\right)}{d} 
+ \lambda_f = 0\;,
\nonumber
\end{eqnarray}
from which we readily derive the condition
\begin{eqnarray}
\frac{1}{d}\sin(\Phi + \theta - \phi_k) = \Lambda\;,
\label{rophi}
\end{eqnarray}
where the Lagrange multipliers were replaced by the more convenient variables $\theta = \arctan\lambda_g$ and 
$\Lambda = -\lambda_f\cos\theta\,$. These variables, as well as condition (\ref{rophi}), can be easily 
interpreted in terms of a geometric construction on the complex plane, as shown in Fig.(\ref{fig:fasores-max-entang}). 
Consider the numbers $e^{i\phi_k}/d$ ($1\leq k\leq d$) rotating in the complex plane constrained by conditions 
(\ref{phid}) and (\ref{pphi}). The overlap is maximized when $d-1$ among these numbers turn around the complex plane 
with the same phase $\phi(t)$ and the remaining one turns in the opposite sense with phase $(1-d)\,\phi(t)\,$. Of course, 
there is an intrinsic permutation symmetry with respect to which one of the eigenvalues will follow the 
opposite evolution, so we may choose the solution $\phi_k = \phi$ $\forall$ $k \in [1,d-1]\,$, while 
$\phi_d = (1-d)\,\phi\,$. This permutation symmetry is broken in the case of non maximal entanglement. 
The geometric interpretation of $\theta$ and $\Lambda$ becomes clear when we represent the line connecting 
the numbers $e^{i\phi_{k}}/d\,$, as depicted in Fig. (\ref{fig:fasores-max-entang}) for the $SU(3)$ case. 
The distance between this connecting line and the origin of the complex plane is $\Lambda\,$, while $\theta$ is 
the angle between this line and the graphic representation of the overlap $O(t)\,$.

It will be easier to express our analytical results in parametrized form as functions of $\phi\,$. 
From the geometric construction of Fig. (\ref{fig:fasores-max-entang}), the following expressions 
can be deduced
\begin{eqnarray}
R_{max} &=& \sqrt{1-4\left(\frac{d-1}{d^2}\right)\sin^2\left(\frac{d\,\phi}{2}\right)}\;,
\nonumber\\
\Phi &=& \phi + \arg{\left(d - 1 + e^{-id\phi}\right)}\;,
\label{analytical}\\
\theta &=& \frac{\pi}{2} - \Phi - \frac{d-2}{2}\,\phi\;,
\nonumber\\
\Lambda &=& \frac{1}{d}\,\cos\left(\frac{d\,\phi}{2}\right)\;.
\nonumber
\end{eqnarray}
Note that the maximum overlap reaches unity when $\phi=2n\pi/d\,$, the allowed topological 
phases for two-qudit systems evolving under local $SU(d)$ operations. The minimum 
value of $R_{max}$ is $1-2/d$ for $\phi=(2n+1)\pi/d\,$. By varying $\phi$ in the interval 
$[0,2\pi]\,$, we can draw the polar plot of the overlap boundary in the complex plane. 
This boundary defines a closed curve with exactly $d$ branches covered by the intervals 
$\phi\in [2n\pi/d,2(n+1)\pi/d]$ ($0\leq n\leq d-1$). For maximally entangled qubits, the 
overlap boundary collapses to a line segment on the real axis defined by the interval 
$[-1,1]\,$. 

\begin{figure}[h!]
\includegraphics[scale=0.48]{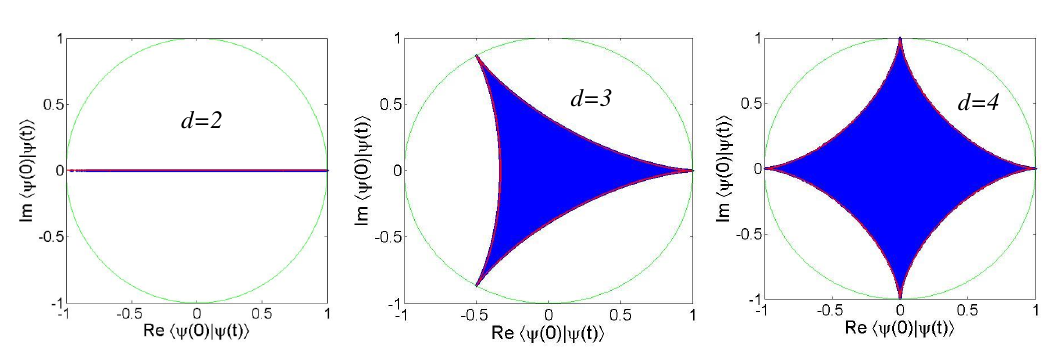}
\caption{\label{fig:random-su-2-3-4} Evolution of the two-qudit overlap under random local 
	$SU(d)$ operations for $d=2,3$ and $4\,$. The analytical boundary is displayed in red 
	(online) and the results of numerical simulations are displayed in blue (online) dots. 
	The unit circle is indicated in green (online) for reference.}
\end{figure}

In order to illustrate the overlap boundaries for different dimensions, we present in 
Fig. (\ref{fig:random-su-2-3-4}) the analytical contours given by Eqs. (\ref{analytical}) 
(red  line curves) and the numerical simulations with random $SU(d)\times SU(d)$ 
transformations (blue  line dots) for $d=2,3$ and $4\,$. 
The numerical results show clear confinement within the analytical boundary.

Following the same geometric construction, the inclusion of partial entanglement is 
straightforward but lengthy and we shall leave the detailed description to a future 
contribution. As a universal behavior in any dimension, the sharp edges between the 
boundary branches are smoothened as entanglement diminishes until the boundary degenerates 
to the unit circle when product states are reached. For qubits, this results in  
inflation of the line segment on the real axis to an ovoid and finally to the 
unit circle. For a pair of qubits initially prepared in a pure state 
with concurrence $C\,$, the solution for the maximum overlap is
\begin{eqnarray}
R_{max} &=& \sqrt{1 - C^2\sin^2\phi}\;,
\nonumber\\
\Phi &=& \arctan\left(\sqrt{1-C^2}\,\tan\phi\right)\;.
\label{solutionsqubits}
\end{eqnarray}
For product states ($C=0$), one trivially obtains $R_{max} = 1$ and $\Phi = \phi\,$, what corresponds 
to the unit circle boundary. For maximally entangled states ($C=1$), we have $R_{max} = \abs{\cos\phi}\in [0,1]$ 
and $\Phi$ becomes discrete, assuming only $0$ or $\pi\,$, the two-qubit topological phases predicted by 
Milman and Mosseri \cite{milman,milman2}. 

\begin{figure}[h!]
	\includegraphics[scale=0.26]{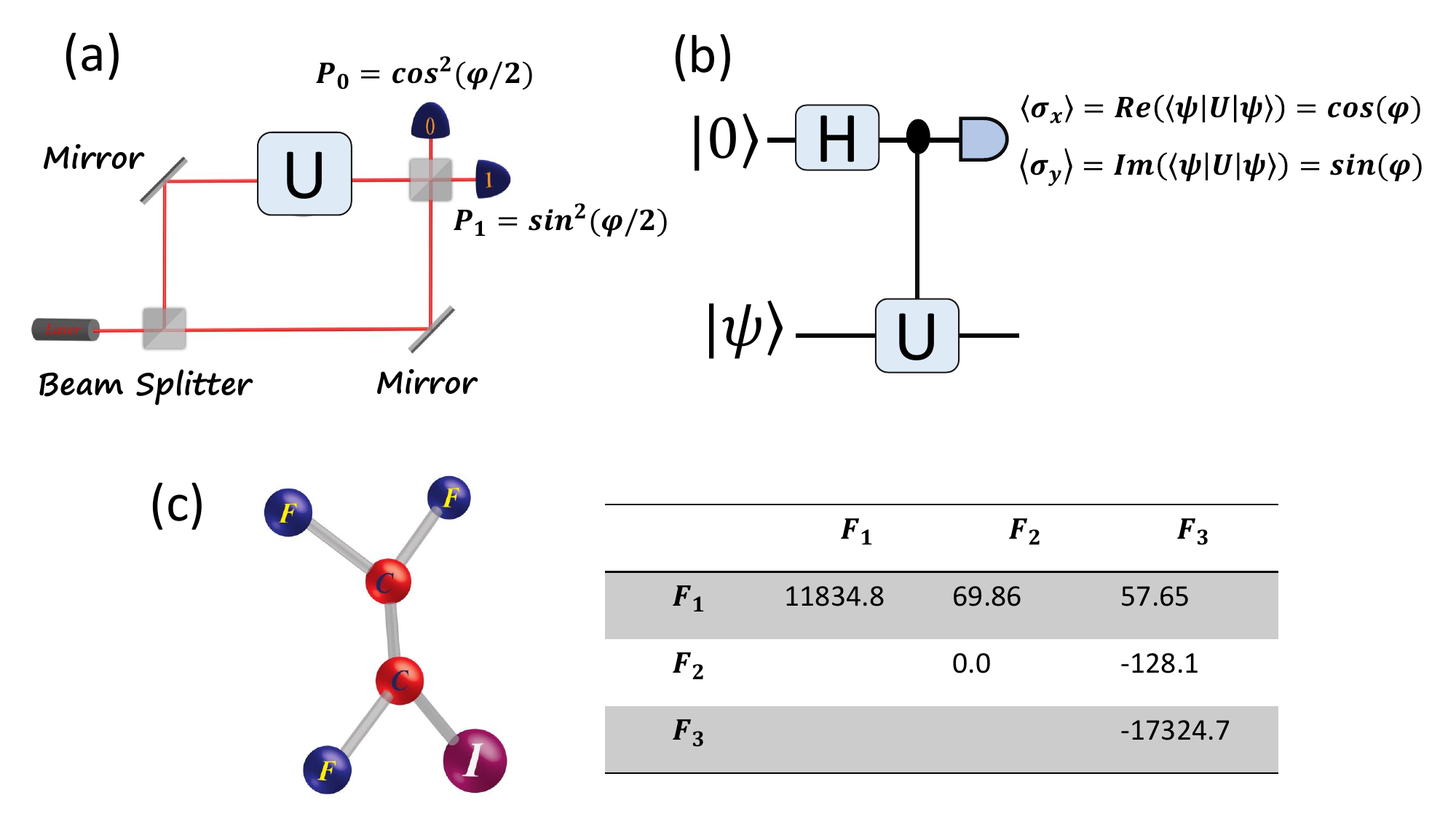}
	\caption{\label{circ} 
		(a) Scheme of an optical interferometer. The phase shift introduced by the unitary $U$ can be determined measuring the probabilities, $P_0$ and $P_1$. 
		(b) The quantum circuit analog to the interferometer based on one auxiliary qubit. The phase shift on the system qubit is determined measuring the  average value of the Pauli matrices, $\sigma_x$ and $\sigma_y$, of the auxiliary qubit. 
		(c) The structure and parameters of the NMR 3-qubit Iodotrifluoroethylene molecule. The diagonal terms in the table are the chemical shifts written in relation to the fluorine nucleus labeled as $F_2$. The off diagonal terms are the coupling constants, all parameters are shown in Hz.}
\end{figure}

Experimental observations of geometrical phases are only achievable using interferometric approaches. Figure \ref{circ}a illustrates the scheme of a simple interferometer. Photons entering the interferometer are split into two perpendicular paths labeled as $|0\rangle$ and $|1\rangle$. On the path $|1\rangle$ the photons undergo a phase shift due to the application of an unitary evolution $U$. After the paths are recombined using the beam splitter the probability for measuring the states $|0\rangle$ or $|1\rangle$ contains information about the phase shift between the paths.

\begin{figure*}[t]

	\begin{subfigure}
		\centering
		\begin{tabular}{c}
			\large{\bf (a)} \\	\large{\bf Target} \\ \large{\bf Concurrence C = 0}\\
			\includegraphics[scale=0.35]{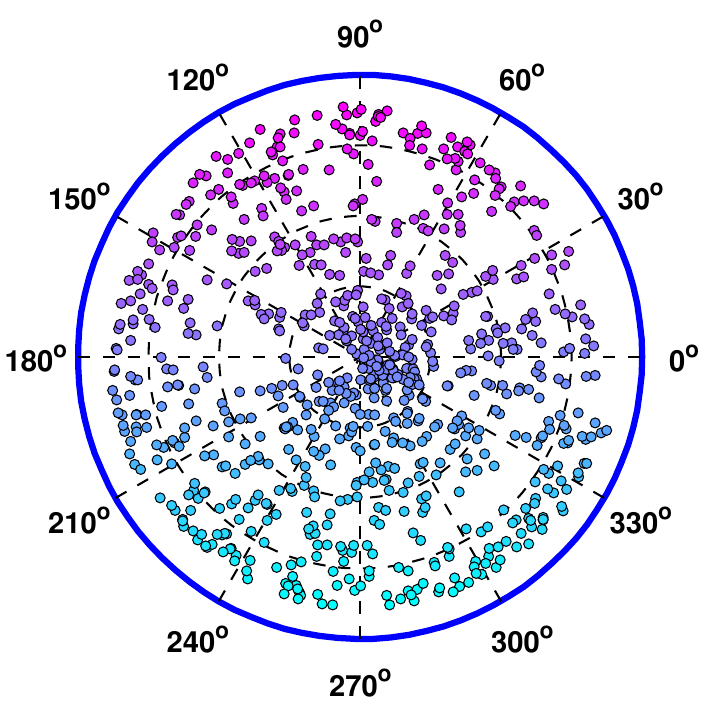}\\
			\includegraphics[scale=0.30]{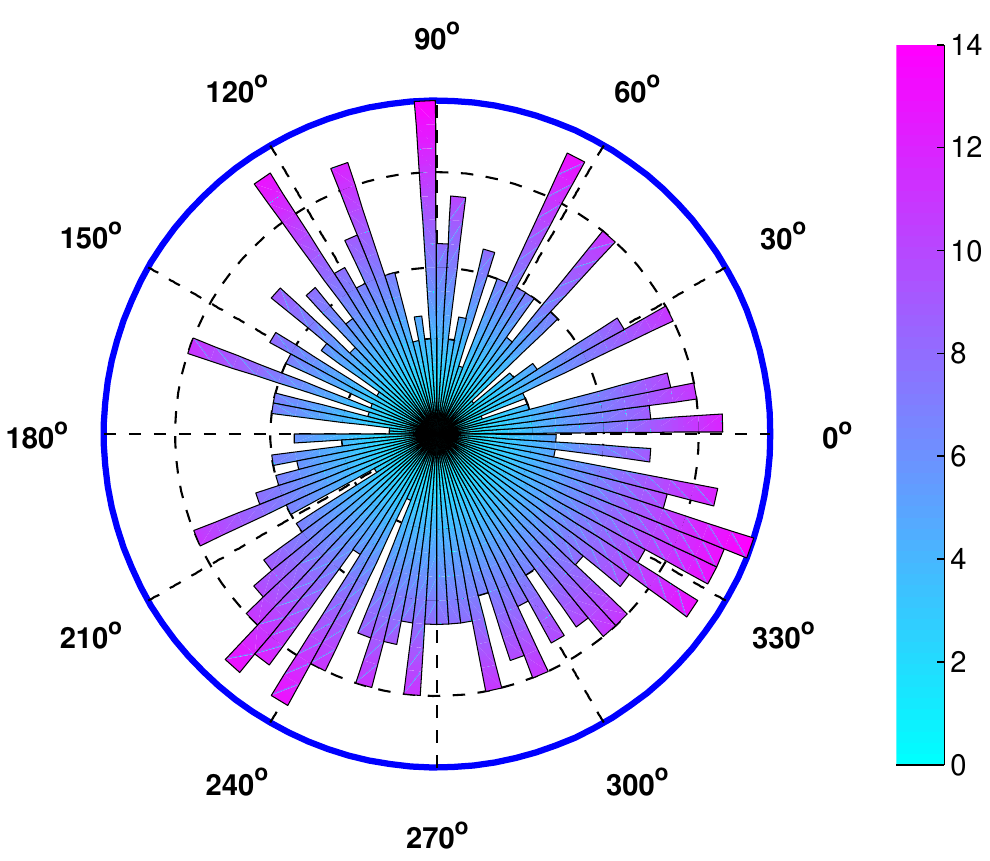}
		\end{tabular}
	\end{subfigure}
	\begin{subfigure}
		\centering
		\begin{tabular}{c}
			\large{\bf (b)} \\	\large{\bf Target} \\ \large{\bf Concurrence C = 0.94}\\
			\includegraphics[scale=0.35]{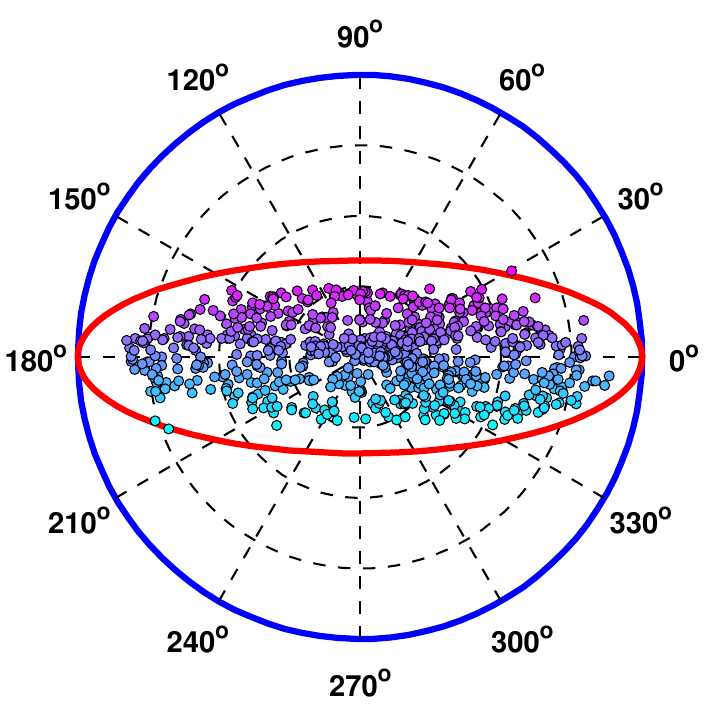}\\
			\includegraphics[scale=0.30]{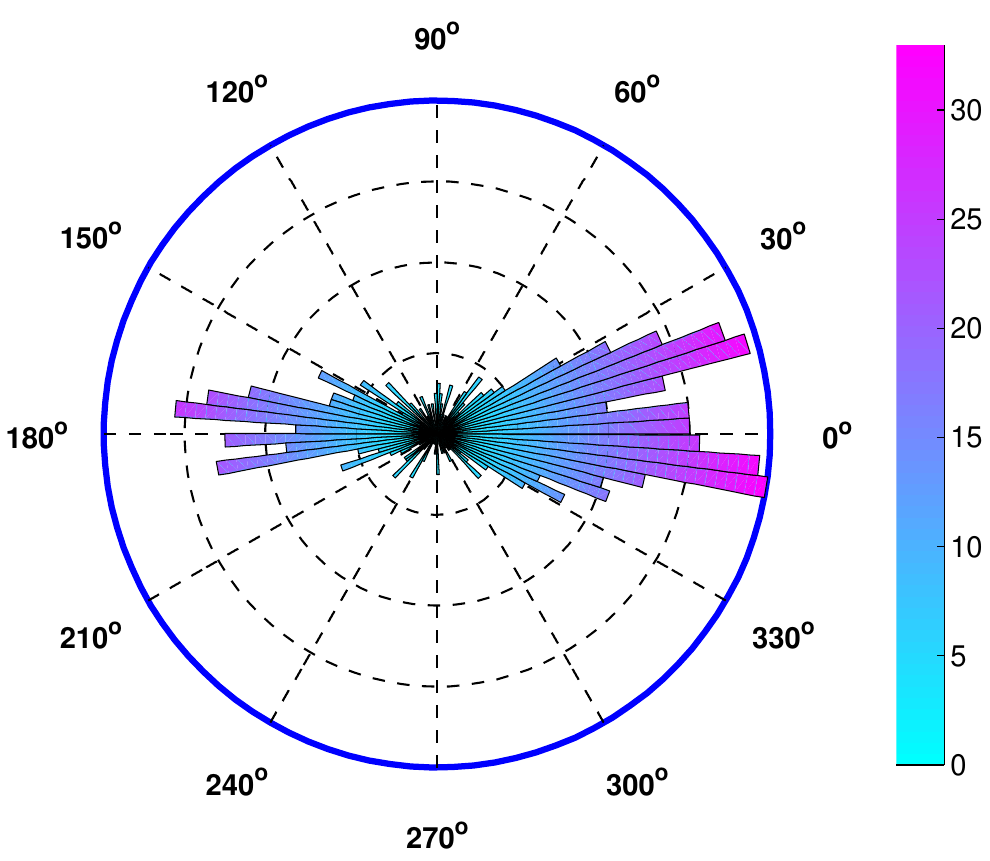}
		\end{tabular}
	\end{subfigure}
	\begin{subfigure}
		\centering
		\begin{tabular}{c}
			\large{\bf (c)} \\	\large{\bf Target} \\ \large{\bf Concurrence C = 1}\\
			\includegraphics[scale=0.35]{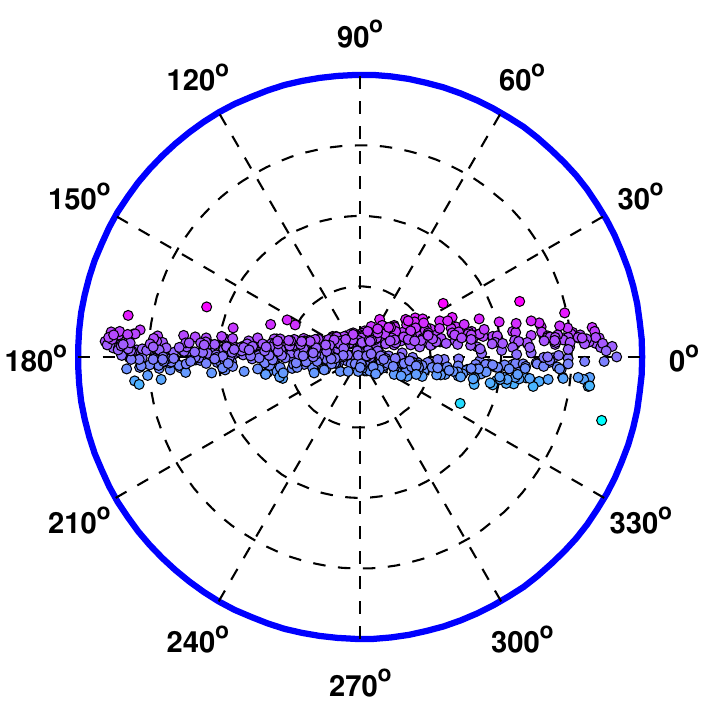}\\
			\includegraphics[scale=0.30]{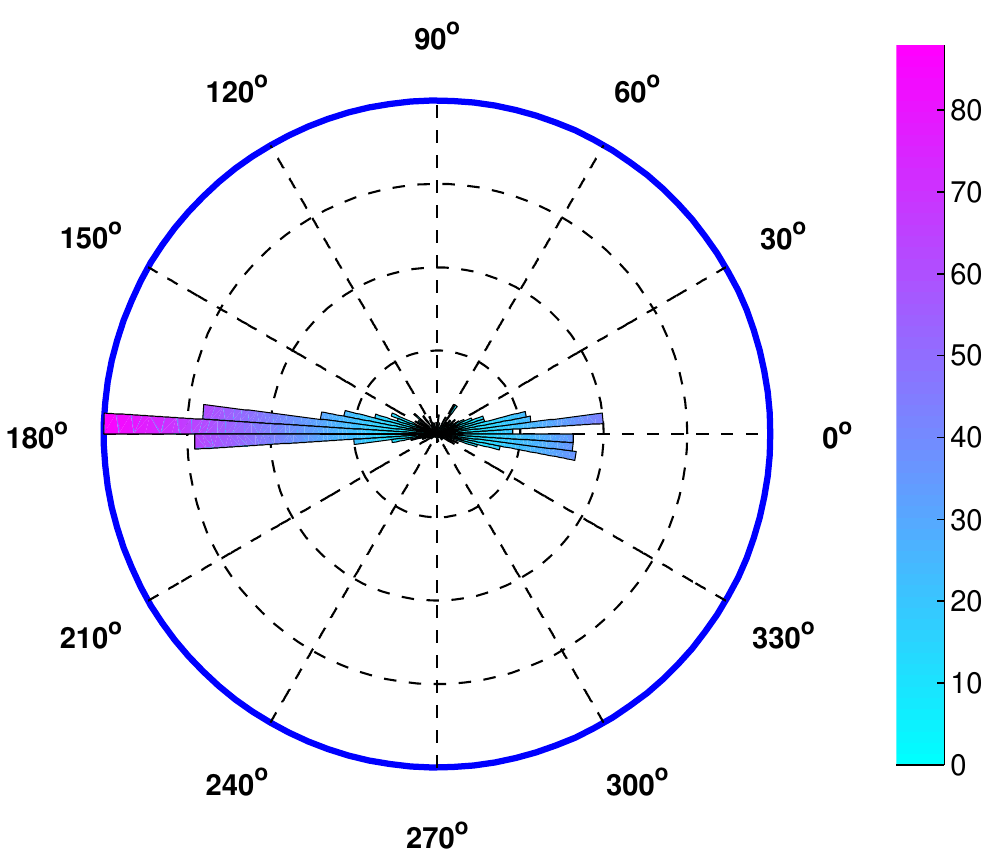}
		\end{tabular}
	\end{subfigure}

	\caption{\label{fig:nmr} Geometrical phase acquired under random unitary evolutions for three different states: 
		(a) The separable state $|00\rangle$, 
		(b) the partial entangled state $\cos(7\pi/36)\,|00\rangle + \sin(7\pi/36)\,|11\rangle$ and 
		(c) the maximally entangled state $(|00\rangle + |11\rangle)/\sqrt{2}$. 
		The top panel shows the overlap $\langle \psi(0)| \psi(t) \rangle$ in the complex plane and the bottom panel shows the distribution of the geometrical phases in polar histograms.}
\end{figure*}

In NMR, the analog of the interferometry model described above can be implemented using the eigenstates of an auxiliary spin$-1/2$ to emulate two photon paths \cite{nmr,suter,kumar,vind} (see figure \ref{circ}b). At the beginning of the interferometry operation the system we want to probe is prepared in a given two-qubit state $|\psi\rangle$, whereas the auxiliary spin is prepared in the superposition state $(|0\rangle + |1\rangle)/\sqrt{2}$, by applying a Hadamard gate on the state $|0\rangle $. The application of the controlled unitary operator $U$ provides the state $(|0\rangle |\psi(0)\rangle+ |1\rangle|\psi(t)\rangle)/\sqrt{2}$, where $|\psi(t)\rangle = U|\psi(0)\rangle = e^{i\phi}|\psi(0)\rangle$ . When a measurement is performed on the auxiliary spin, its normalized $x-y$ magnetization components on the plane, which are proportional to the average value of the Pauli matrices, are directly related to the overlap between the states.

To perform the experiment we have used an ensemble of identical and non-interacting molecules  in liquid state at room temperature, where nuclear spins are employed as qubits. The quantum state of the ensemble is prepared, from the thermal equilibrium, in the so called pseudo-pure state (PPS) \cite{livro}:
\begin{eqnarray}
\rho = (1-\epsilon) \frac{I}{D} + \epsilon |\psi\rangle \langle \psi|.
\label{pps}
\end{eqnarray}

Since the maximally mixed part $I/D$ does not produce an 
observable signal, the overall NMR signal arises only from the pure state part $|\psi\rangle \langle \psi|$ where the factor 
$\epsilon \approx 10^{-5}$ is the thermal polarization and $D$ is the dimension of the Hilbert space. Under a suitable normalization, the state (\ref{pps}) is equivalent to that from a system in a pure state $|\psi\rangle$, and therefore, can be used to test different features of pure entangled states \cite{bell1,bell2,ghz}. 

The experiment was performed at room temperature, using a 500 MHz Varian NMR spectrometer and an ensemble of Iodotrifluoroethylene ($C_2F_3I$) molecules dissolved in deuterated acetone. This molecule (see figure \ref{circ}c) contains three spin-$1/2$ $^{19}$F nuclei, where two 
fluorine nuclei are used to encode the state $|\psi\rangle$ and one fluorine nucleus is used as the auxiliary spin. The PPS was created using the control transfer gates technique \cite{transfer}, the actual pulse sequence used can be found in \cite{vind}. To implement gate operations we exploit standard Isech shaped pulses interleaved with periods of free evolutions. For combining all operations into a single pulse sequence we have used the techniques described in \cite{laflamme1,laflamme2,laflamme3}.  

Figure \ref{fig:nmr} shows the overlap  $\langle \psi(0)| \psi(t) \rangle$ in the complex plane under 800 random unitaries for three states with different degrees of entanglement. The unitaries were applied only on a single qubit of the entangled pair and have the form $U = R_x(\theta) R_z(\beta)$, where $\theta$ and $\beta$ were randomly chosen from $0$ to $4 \pi$. The results clearly shows the confinement by the bounds (\ref{solutionsqubits}). As the degree of entanglement is increased the acquired geometrical phase become more robust, being only  $\pm \pi$ when the state is maximally entangled. The deviations from the theory are due to decoherence processes and experimental imperfections on the control gates.

In conclusion, we have investigated pairs of entangled qudits evolving under random, local $SU(d)$ operations. More specifically we have analytically demonstrated that the overlap between the $SU(d)$ transformed and the initial state is confined within boundaries that only depend on the degree of entanglement. This feature was illustrated with numerical simulations and, moreover, experimentally verified with an NMR setup. Our results provide a step further in the direction of robust quantum computation and may find application for fault tolerant implementations of phase gates. They also open new perspectives for investigations of nontrivial bounds on multi-qubit systems, where fractional phases are expected as well \cite{multiqubit,multiqubit2}.

\section*{Acknowledgments}

Funding was provided by Coordena\c c\~{a}o de Aperfei\c coamento de 
Pessoal de N\'\i vel Superior (CAPES), Funda\c c\~{a}o de Amparo \`{a} 
Pesquisa do Estado do Rio de Janeiro (FAPERJ-BR), and Instituto Nacional 
de Ci\^encia e Tecnologia de Informa\c c\~ao Qu\^antica (INCT-CNPq).


\begin{thebibliography}{99}

%%%%%%%%%%%%%%%%%%%%%%%%%%%% Reservoir engineering %%%%%%%%%%%%%%%%%%%%%%%%%%%%%%%%%%%%%
%
\bibitem{kwiat}
N. A. Peters, J. B. Altepeter, D. A. Branning, E. R. Jeffrey, T.-C. Wei, P. G. Kwiat, 
\prl \textbf{92}, 133601 (2004).
%
\bibitem{aiello}
A. Aiello, G. Puentes, and J. P. Woerdman, 
\pra \textbf{76}, 032323 (2007).
%
\bibitem{global}
M. Hor-Meyll, A. Auyuanet, C. V. S. Borges, A. Arag\~ao, J. A. O. Huguenin, A. Z. Khoury, L. Davidovich,
% Environment-induced entanglement with a single photon
\pra \textbf{80}, 042327 (2009).

%%%%%%%%%%%%%%%%%%%%%%%%%%%% Decoherence free subspaces %%%%%%%%%%%%%%%%%%%%%%%%%%%%%%%%%%%%%
%
\bibitem{decfree}
G. M. Palma, K. A. Suominen, A. K. Ekert, 
Proc. R. Soc. London, Ser. A \textbf{452}, 567 (1996).

%%%%%%%%%%%%%%%%%%%%%%%%%%%% Geometric phases computation %%%%%%%%%%%%%%%%%%%%%%%%%%%%%%%%%%%%%

%
\bibitem{vedral}
J. A. Jones, V. Vedral, A. Ekert, and G. Castagnoli,
%"Geometric quantum computation using nuclear magnetic resonance," 
Nature (London) {\textbf 403}, 869 (2000).
%

%%%%%%%%%%%%%%%%%%%%%%%%%%%% Topological quantum information %%%%%%%%%%%%%%%%%%%%%%%%%%%%%%%%%%%%%

%
\bibitem{faulttolerant} 
A. Kitaev, Ann. Phys. (N.Y.) \textbf{303}, 2 (2003).

%%%%%%%%%%%%%%%%%%%%%%%%%%%% Geometric phase entanglement %%%%%%%%%%%%%%%%%%%%%%%%%%%%%%%%%%%%%%%%%


\bibitem{dd1}
% Protected Quantum Computing: Interleaving Gate Operations with Dynamical Decoupling Sequences
J. Zhang, A. M. Souza, F. D. Brandao and D. Suter  
{\it Phys. Rev. Lett.}  {\bf 112}, 050502 (2014).


\bibitem{dd2}
% Robust Dynamical Decoupling for Quantum Computing and Quantum Memory
A. M. Souza, G. A. Alvarez, and D. Suter
{\it Phys. Rev. Lett.}  {\bf 106}, 240501 (2011).

\bibitem{dd3}
% Experimental protection of quantum gates against decoherence and control errors
A. M. Souza, G. A. Alvarez, and D. Suter
{\it Phys. Rev. A}  {\bf 86}, 050301(R) (2012).


\bibitem{dd4}
% High-fidelity gate operations for quantum computing beyond dephasing time limits
A. M. Souza, R. S. Sarthour, I. S. Oliveira, and D. Suter
{\it Phys. Rev. A}  {\bf 96}, 062332 (2015).


%
\bibitem{sjoqvist}
E. Sj\"oqvist, 
\pra \textbf{62}, 022109 (2000).
%
\bibitem{sjoqvist2}
B. Hessmo and E. Sj\"oqvist, 
\pra \textbf{62}, 062301 (2000).
%

%%%%%%%%%%%%%%%%%%%%%%%%%%%% Geometric phase entanglement qubits 0/Pi %%%%%%%%%%%%%%%%%%%%%%%%%%%%%%%%%%%%%%%%%

%
\bibitem{milman} 
P. Milman, and R. Mosseri, 
\prl \textbf{90}, 230403 (2003).
%
\bibitem{milman2} 
P. Milman, \pra \textbf{73}, 062118 (2006).
%
\bibitem{topoluff}
C. E. R. Souza, J. A. O. Huguenin, P. Milman, and A. Z. Khoury, 
%Topological phase for spin-orbit transformations on a laser beam,
\prl \textbf{99}, 160401 (2007). 
%
\bibitem{nmr}
%Experimental observation of a topological phase in the maximally entangled state of a pair of qubits
J. Du, J. Zhu, M. Shi, X. Peng, D. Suter, 
\pra \textbf{76}, 042121 (2007).
%

%%%%%%%%%%%%%%%%%%%%%%%%%%%% Geometric phase entanglement qudits 0-2Pi/d %%%%%%%%%%%%%%%%%%%%%%%%%%%%%%%%%%%%%%%%%

%
\bibitem{fracuff}
%Fractional Topological Phase for Entangled Qudits
L. E. Oxman and A. Z. Khoury,
\prl {\bf 106}, 240503 (2011).
%
\bibitem{fracuffufmg}
%Fractional topological phase on spatially encoded photonic qudits
A. Z. Khoury, L. E. Oxman, B. Marques, A. Matoso, and S. P\'adua, 
\pra {\bf 87}, 042113 (2013).
%
\bibitem{fracuff2}
%Topological phase structure of entangled qudits
A. Z. Khoury and L. E. Oxman,
\pra {\bf 89}, 032106 (2014).
%
\bibitem{fracuffufmg2}
%Experimental observation of fractional topological phases with photonic qudits
A. A. Matoso, X. S\'anchez-Lozano, W. M. Pimenta, P. Machado, B. Marques, F. Sciarrino, 
L. E. Oxman, A. Z. Khoury, S. P\'adua,
%Fractional topological phase on spatially encoded photonic qudits
\pra \textbf{94}, 052305 (2016).
%

%%%%%%%%%%%%%%%%%%%%%%%%%%%%%%%% Multiple qubit geometric phase %%%%%%%%%%%%%%%%%%%%%%%%%%%%%%%%

%
\bibitem{multiqubit}
%Topological phases and multiqubit entanglement
M. Johansson, M. Ericsson, K. Singh, E. Sj\"oqvist, and M. S. Williamson,
\pra {\bf 85}, 032112 (2012).
%
\bibitem{multiqubit2}
%Three-qubit topological phase on entangled photon pairs
M. Johansson, A. Z. Khoury, K. Singh, and E. Sj\"oqvist, 
\pra {\bf 87}, 042112 (2013).


%%%%%%%%%%%%%%%%%%%%%%%%%%%%%%%% Qudit quantum computation %%%%%%%%%%%%%%%%%%%%%%%%%%%%%%%%

%
\bibitem{bullock}
% Asymptotically Optimal Quantum Circuits for d-Level Systems
S. S. Bullock, D. P. O'Leary, and G. K. Brennen, 
\prl {\bf 94}, 230502 (2005).
%
\bibitem{ashok}
% Multivalued logic gates for quantum computation 
A. Muthukrishnan, and C. R. Stroud, Jr., 
\pra {\bf 62}, 052309 (2000).
%
\bibitem{munro}
% Generalized Toffoli gates using qudit catalysis
R. Ionicioiu, T. P. Spiller, and W. J. Munro, 
\pra {\bf 80}, 012312 (2009).
%


\bibitem{suter}
% Study of the Aharonov-Anandan Quantum Phase by NMR Interferometry
D. Suter, K. T. Mueller, and A. Pines, 
\prl {\bf 60}, 1218 (1998).

\bibitem{kumar}
% Experimental measurement of mixed state geometric phase
%by quantum interferometry using NMR
A. Ghosh and A. Kumar, 
{\it Physics Letters A} {\bf 349}, 27-36 (2006).

\bibitem{vind}
% Experimental realization of the Yang-Baxter Equation via NMR interferometry
F. Anvari Vind, A. Foerster, I. S. Oliveira, R. S. Sarthour, D. O. Soares-Pinto, A. M. Souza and I. Roditi,
{\it Scientific Reports} {\bf 6}, 20789 (2016).

\bibitem{laflamme1}
% Compiling gate networks on an Ising quantum computer
M. D. Bowdrey, J. A. Jones, E. Knill, and R. Laflamme, 
\pra {\bf 72}, 032315 (2005).

\bibitem{laflamme2}
% Liquid-state nuclear magnetic resonance as a testbed for developing quantum control methods
C. A. Ryan, C. Negrevergne, M. Laforest, E. Knill, and R. Laflamme, 
\pra {\bf 78}, 012328 (2008).

\bibitem{laflamme3}
% Experimental magic state distillation for fault-tolerant quantum computing
A. M. Souza, J. Zhang, C. A. Ryan, and R. Laflamme, 
{\it Nature Communications}  {\bf 2169}, 1 (2011).

\bibitem{livro}
I.S.Oliveira, T.J. Bonagamba,R. S. Sarthour, J. C. C. Freitas and E. R. deAzevedo {\it NMR Quantum Information Processing (Amsterdam: Elsevier)} (2007).


\bibitem{bell1}
% NMR analog of Bell's inequalities violation test.
 A. M. Souza, A. Magalhães, J. Teles, E. R. deAzevedo, T. J. Bonagamba, I. S. Oliveira and R. S. Sarthour,  
{\it New J. Phys.}  {\bf 10}, 033020 (2008).

\bibitem{bell2}
% Experimentally simulating the violation of Bell-type inequalities for generalized GHZ states.
Ren, Changliang; Lu, Dawei; Shi, Mingjun; et al.,  
{\it Phys. Lett. A}  {\bf 373}, 4222 (2009).

\bibitem{ghz}
% experimental demonstration of Greenberger-Horne-Zeilinger correlations using nuclear magnetic resonance
 Richard J. Nelson, David G. Cory, and Seth Lloyd,  
{\it Phys. Rev. A}  {\bf 61}, 022106 (2000).


\bibitem{transfer}
% Preparing pseudopure states with controlled-transfer gates
M. Kawamura, B. Rowland and J. A. Jones
{\it Phys. Rev. A}  {\bf 82}, 032315 (2010).


\end{thebibliography}
\end{document}